\documentclass[preprint,prl,aps,showpacs]{revtex4}
\usepackage{graphicx}
\usepackage{dcolumn}
\usepackage{bm}
\usepackage{amsmath}
\begin{document}
\title{First-principles Predictor of the Location of Ergodic-Non-ergodic Transitions}
\author{P. E. Ram\'{\i}rez-Gonz\'alez$^{1}$, R.
Ju\'arez-Maldonado$^{1}$, L. Yeomans-Reyna$^{2}$, M. A.
Ch\'avez-Rojo$^{3}$,  M. Ch\'avez-P\'aez$^{1}$, A.
Vizcarra-Rend\'on$^{4}$, and M. Medina-Noyola$^1$}

\address{\\$^{1}$Instituto de F\'{\i}sica {\sl ``Manuel Sandoval Vallarta"},
Universidad Aut\'{o}noma de San Luis Potos\'{\i}, \'{A}lvaro
Obreg\'{o}n 64, 78000 San Luis Potos\'{\i}, SLP, M\'{e}xico\\
$^{2}$Departamento de F\'{\i}sica, Universidad de Sonora, Boulevard Luis\\
Encinas y Rosales, 83000, Hermosillo, Sonora, M\'{e}xico.\\
$^{3}$Facultad de Ciencias Qu\'{\i}micas, Universidad Aut\'onoma de
Chihuahua, Venustiano Carranza S/N, 31000 Chihuahua, Chih.,
M\'exico.\\$^{4}$Unidad Acad\'emica de F\'{\i}sica, Universidad
Aut\'onoma de Zacatecas, Paseo la Bufa y Calzada Solidaridad, 98600,
Zacatecas, Zac.,  M\'{e}xico. }
\date{\today}

\begin{abstract}
This letter presents a remarkably simple approach to the
first-principles determination of the ergodic-non-ergodic transition
in monodisperse colloidal suspensions. It consists of an equation
for the long-time asymptotic value $\gamma$ of the mean squared
displacement of the colloidal particles, whose finite real solutions
signal the non-ergodic state, and determines the non-ergodic
parameter $f(k)$. We illustrate its concrete application to three
simple model colloidal systems, namely, hard-spheres, hard-spheres
plus repulsive (screened Coulomb) Yukawa interaction, and
hard-sphere plus attractive Yukawa tail. The results indicate that
this is quite a competitive theory, similar in spirit to, but
conceptually independent from, the well-known mode coupling theory.


\bigskip

\noindent
Esta carta presenta un m\'etodo notablemente simple para
la determinaci\'on, de primeros principios, de la transici\'on
erg\'odico--no-erg\'odico en suspensiones coloidales monodispersas.
Dicho m\'etodo consiste en una ecuaci\'on para el valor asint\'otico
a tiempos largos, $\gamma$, del desplazamiento cuadr\'atico medio de
las part\'{\i}culas coloidales, cuyas soluciones reales finitas son
sin\'onimo de no-ergodicidad, y determinan el par\'ametro no
erg\'odico $f(k)$. Ilustramos su aplicaci\'on concreta en tres
modelos simples de sistemas coloidales, a saber, esferas duras,
esferas duras con interacci\'on de Yukawa repulsiva (Coulombica
apantallada), y esferas duras con interacci\'on de Yukawa atractiva
(fuerzas de depleci\'on). Los resultados indican que \'esta es una
teor\'{\i}a muy competitiva, similar en esp\'{\i}ritu, pero
conceptualmente diferente, a la bien conocida teor\'{\i}a de {\it
acoplamiento de modos}.

\medskip\noindent Descriptores: din\'amica coloidal, transici\'on
v\'{\i}trea, arresto dinámico.

\end{abstract}

\maketitle

The most basic and elementary information on the thermodynamic
properties of a material is its phase diagram. The description of
the gas-liquid transition provided by the van der Waals equation of
state is the earliest and most paradigmatic example of the
construction of a phase diagram starting from molecular
considerations \cite{mjklein}. The most outstanding achievement of
statistical mechanics, however, has been the establishing of the
microscopic version of the second law of thermodynamics, which
provides the basis for the systematic calculation, given the
intermolecular forces, of phase diagrams, by the simple conceptual
procedure of identifying the phase with the lowest free energy. This
very fundamental principle allowed the development of the wide range
of methods, approaches, techniques and applications of equilibrium
statistical thermodynamics \cite{mcquarrie}.

At the same time, one of the most notorious limitations of
statistical mechanics has been its inability to identify an equally
general and simple principle that allows us to describe
non-equilibrium states of matter, given the molecular interactions.
From a practical and fundamental perspective, this is quite
disturbing, given the fact that a large variety of the materials
with which we actually interact in ordinary life are not in their
thermodynamic equilibrium state. Thus, it is permanently important
to search for first-principles approaches to describe the most
elementary properties of such ``phases" and the transitions between
them, even in the context of specific classes of non-equilibrium
states. Perhaps the simplest of them refer to the states that result
when kinetic barriers prevent a material from reaching its
thermodynamically most stable state, thus being trapped in
dynamically arrested states, such as glasses or gels. One then would
like to have first-principles criteria to predict the location of
the boundary between the ergodic fluid phase and such arrested
non-ergodic states. Contrary to its thermodynamic analog, in this
case we only have a single theory of this sort, namely, the mode
coupling theory (MCT) of the ideal glass transition \cite{goetze1,
goetze2}, some of whose predictions have found beautiful
experimental confirmation \cite{vanmegen2, beck}. Unfortunately, its
criterion to decide in a practical manner if a system is in an
ergodic or in an arrested state is somewhat obscured by the
conceptual complexity of this theory and by the many other issues
addressed \cite{goetze1}. Thus, the need exists of simpler and more
straightforward approaches that focus on this specific and important
issue. The main purpose of this communication is to present a new
first-principles prescription to locate the ergodic-non-ergodic
transition in monodisperse colloidal dispersions in a remarkably
practical and simple manner, completely independent of the MCT and
its recent variants \cite{szamelmc, cao}.

This prescription consists of an equation for the most elementary
``order parameter" associated to the transition from an equilibrium
fluid phase to a glass or gel state. This is the long-time
asymptotic value, $\gamma$, of the mean squared displacement (msd)
of the particles in the suspension. In the arrested states, this
parameter is finite, representing the localization of the particles,
whereas in the ergodic phase, it diverges. Such an equation reads

\begin{equation}
\frac{1}{6\pi^{2}n}\int_{0}^{\infty } dkk^4\frac{\left[S(k)-1\right]
^{2}\lambda^2 (k)\gamma}{\left[\lambda (k)S(k) +
k^2\gamma\right]\left[\lambda (k) + k^2\gamma\right]}=1,
\label{nep5pp}
\end{equation}

\noindent where $S(k)$ is the static structure factor of the system,
$\lambda (k)\equiv [1+(k/k_c)^2]^{-1}$, with $k_c$ being the
position of the first minimum of $S(k)$, and $n$ is the particle
number concentration. The very form of this criterion exhibits its
simplicity: given the effective inter-particle forces, statistical
thermodynamic methods allows one to determine $S(k)$, and the
absence or existence of real solutions of this equation will
indicate if the system remains in the ergodic phase or not. The next
objective of this letter is to illustrate the practical application
of this criterion, in the context of three simple model colloidal
systems, and to explain very briefly its conceptual origin.

\begin{figure}
\includegraphics[scale=.28]{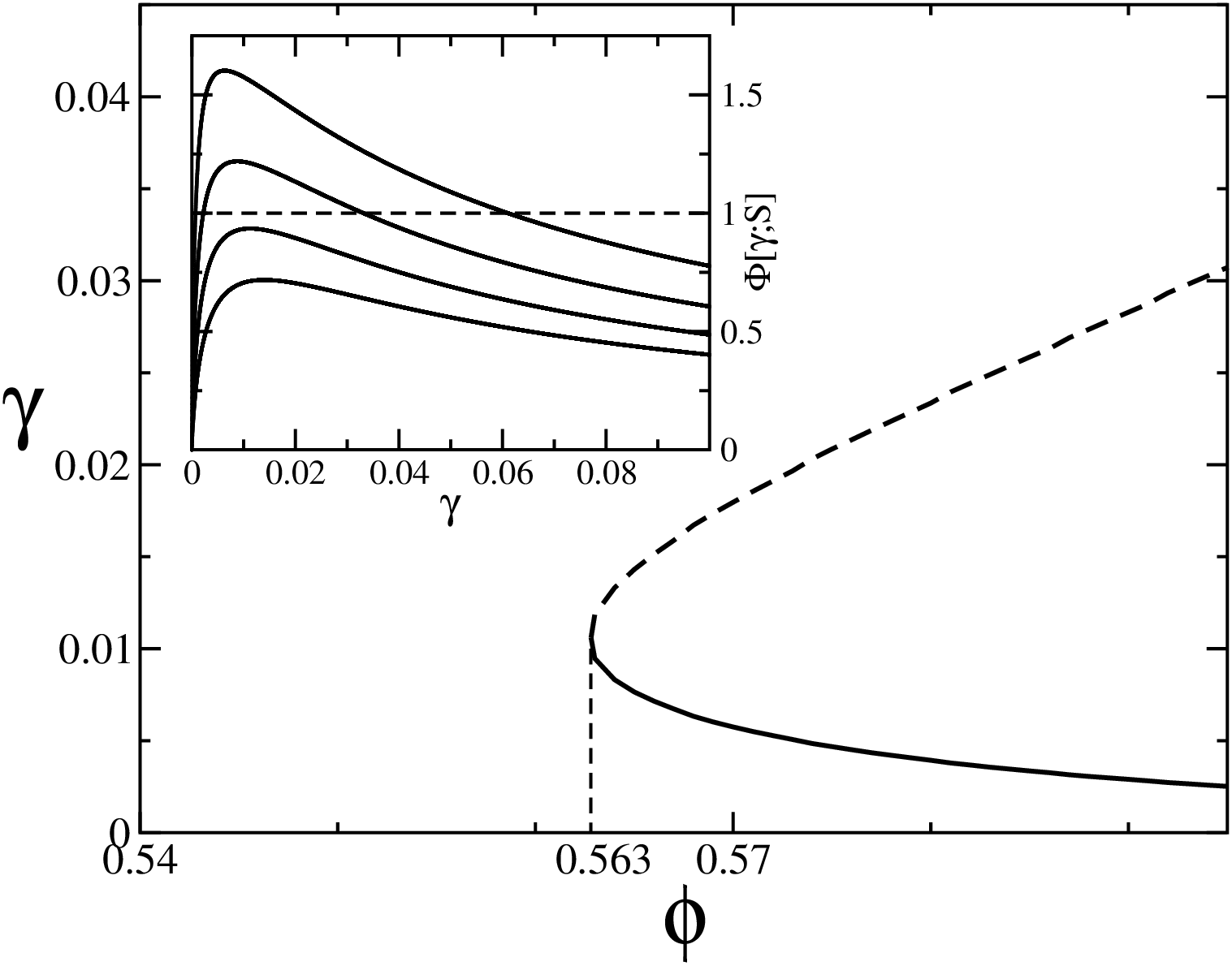}
\caption{Real solutions $\gamma$ of Eq.\ (\ref{nep5pp}). Below
$\phi_g=0.563$, this equation has no real solutions. Above $\phi_g$,
two solutions appear, illustrated by the two branches that bifurcate
at $\phi_g$. The branch for which $\gamma$ decreases with $\phi$
(solid line) corresponds to the physical solution of the glass
state. In the inset we plot the functional
$\Phi\left[\gamma;S\right]$ as a function of $\gamma$ for $\phi=
0.50, 0.55$, $0.60$, and $0.65$ (from bottom to top).} \label{fig.1}
\end{figure}

To illustrate the use of this criterium, let us notice that the
left-hand side of Eq.\ (\ref{nep5pp}) is a functional of $S(k)$ and
an ordinary function of $\gamma$, which we denote by
$\Phi\left[\gamma;S\right]$. Thus, for a fixed state [i.e., fixed
$S(k)$], this equation  may be solved by plotting
$\Phi\left[\gamma;S\right]$ as a function of $\gamma$, to see if it
crosses unity, and for which value(s) of $\gamma$ it does so; notice
that this functional must vanish in the limits of large and small
$\gamma$. This very simple procedure leads to the determination
of the finite solutions of Eq.\ (\ref{nep5pp}), as illustrated in
Fig.\ \ref{fig.1} for the hard sphere (HS) system with the static
structure factor $S(k)$ given by the Percus-Yevick
(PY) approximation \cite{percusyevick} with the Verlet-Weis (VW)
correction \cite{verletweis}. The inset of Fig.\ \ref{fig.1}
exhibits the dependence of the functional on $\gamma$ for various
volume fractions. Clearly, below a threshold value $\phi _g$,
$\Phi\left[\gamma;S\right]$ remains below $1.0$ for all $\gamma$,
and hence, there are no real solutions; thus, the system must be in
the ergodic state for $\phi < \phi _g$. Above $\phi _g$ there are
two real solutions, the smallest one corresponding to the glass
state, since in the glass the msd must decrease with $n$. In this
manner, we determine this threshold value to be $\phi _g=0.563$.
This quantitative prediction is closer to the experimental values
than those of the original MCT or its extensions.

The criterion above emerges from the long-time asymptotic analysis
\cite{todos1} of the self-consistent generalized Langevin equation
(SCGLE) theory of colloid dynamics \cite{5, TAVSDB}. This theory was
originally developed to describe the dynamic properties of colloidal
dispersions. Thus, it allows the calculation of properties such as
the msd of the particles, or the intermediate scattering function
$F(k,t)$ and its {\it self} component, $F_S(k,t)$ \cite{1,2}, given
the effective pair potential $u(r)$ between colloidal particles.
From the same long-time analysis one can also derive \cite{todos1} a
remarkably simple expression for another experimentally important
property of the glass states, namely, the non-ergodic parameter
$f(k) \equiv F(k,t\to \infty)/S(k)$. Such an expression reads

\begin{equation}
f(k)=\frac{\lambda (k)S(k)}{\lambda (k)S(k) + k^2\gamma},
\label{nep1p}
\end{equation}

\noindent
with $\gamma$ being the physical solution of Eq.\ (\ref{nep5pp}). Let
us notice that this result only involves $S(k)$ and $\lambda(k)$ as
static inputs. We may illustrate its use by applying it to the HS
system above, for which we just found  $\phi _g=0.563$. Right
at $\phi _g$, there is only a single solution for $\gamma$, namely,
$\gamma= 1.06\times 10^{-2}\sigma^2$, with $\sigma$ being the
HS diameter. This solution for $\gamma$ may then be
employed in Eq.\ (\ref{nep1p}) to determine the non-ergodic parameter
$f(k)$. Fig. 2 compares this prediction for $f(k)$ with the
experimental data of Ref.\ \cite{vanmegen2}. A similar comparison is
also included in this figure, corresponding to a higher volume
fraction, $\phi=0.58$, and to the experimental data of
Ref.\ \cite{beck}.

\begin{figure}
\includegraphics[scale=.28]{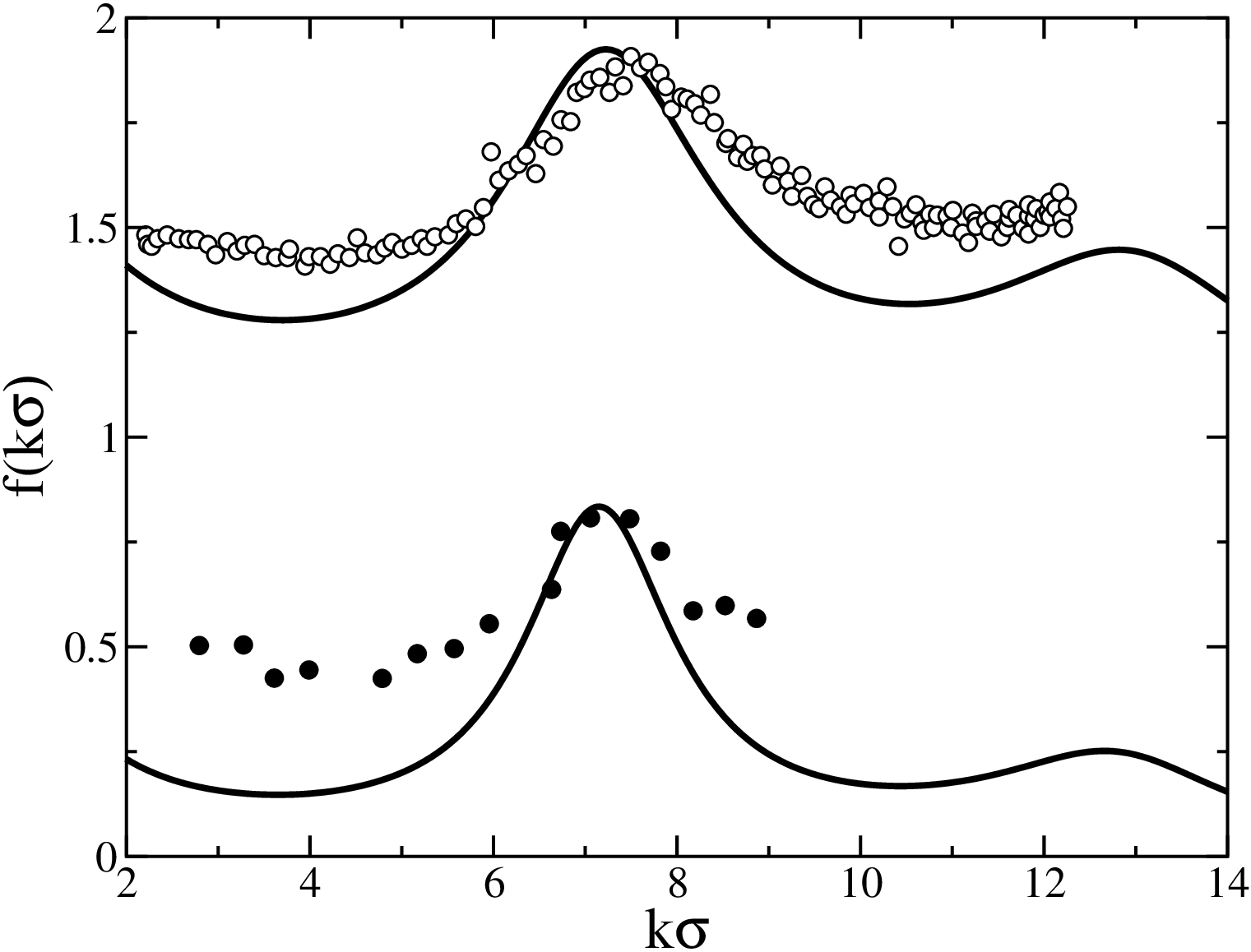}
\caption{Non-ergodic parameter $f(k)$ (solid lines) calculated with
Eqs.\ (\ref{nep5pp}) and (\ref{nep1p}) for the HS system,
with $S(k)$ given by the PY approximation with the
VW correction, at the theoretically-predicted ideal glass
transition, $\phi _g=0.563$ (lower curve), and at the larger volume
fraction $\phi=0.58$ (upper curve; shifted upwards one unit, for
visual clarity). Symbols are the experimental data of
Ref.\ \cite{vanmegen2} (filled circles) and Ref.\ \cite{beck}
(empty circles). }
\label{fig.4}
\end{figure}

The next example refers to dispersions of charged colloidal
particles, with effective pair potential modeled by the HS plus the
repulsive screened Coulomb potential, $\beta u(r)=K\exp[-z(r/\sigma
-1)]/(r/\sigma )$ for $r>\sigma$. The interaction parameters $z$ and
$K$ are, respectively, the inverse Debye screening length (in units
of $\sigma$), and the intensity of the pair potential at hard-sphere
contact (in units of the thermal energy  $\beta^{-1}\equiv k_BT$).
We employed as the static input of Eqs.\ (\ref{nep5pp}) and
(\ref{nep1p}) the experimentally-determined static structure factor
of a sample with $\sigma= 272 nm$ and $\phi=0.27$, provided in Ref.\
\cite{beck}. For this, we used the hyper-netted chain
approximation \cite{mcquarrie} to provide a smooth fit of the data,
leading to $z=3.14$ and $K=11.56$. The solution of Eq.\
(\ref{nep5pp}) is $\gamma= 0.00293\sigma^2$, and the results for
$f(k)$ from Eq.\ (\ref{nep1p}) are compared with the corresponding
experimental data in Fig. 3. Given that no fitting parameters are
involved in the theoretically-predicted non-ergodic parameter $f(k)$
in Figs. 2 and 3, we consider that the overall quantitative accuracy
of our theory is quite remarkable, and is certainly better than that
of the MCT. Of course, one can easily determine the full
liquid-glass ``phase" diagram in the space ($K, z, \phi$), but this
is not discussed here.

\begin{figure}
\begin{center}
\includegraphics[scale=.28]{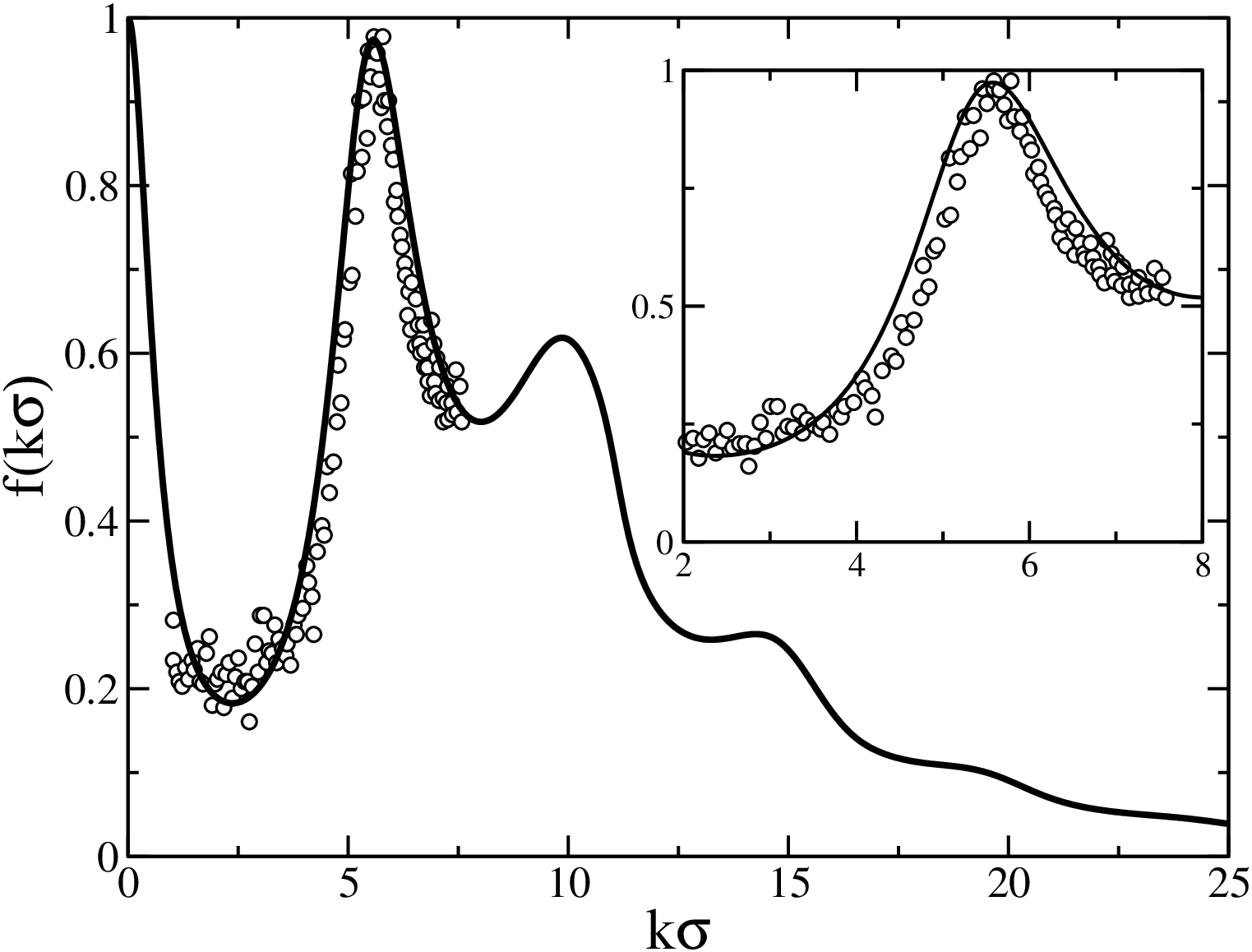}
\caption{Theoretical prediction of the non-ergodic parameter $f(k)$
(solid line) calculated with Eqs.\ (\ref{nep5pp}) and (\ref{nep1p})
for the hard-sphere plus the {\it screened} Coulomb tail, with
$\phi=0.27$, $z=3.14 $, and $K=11.56$. The symbols are the
experimental data of Ref.\ \cite{beck}} \label{fig.6}
\end{center}
\end{figure}

\begin{figure}
\begin{center}
\includegraphics[scale=.28]{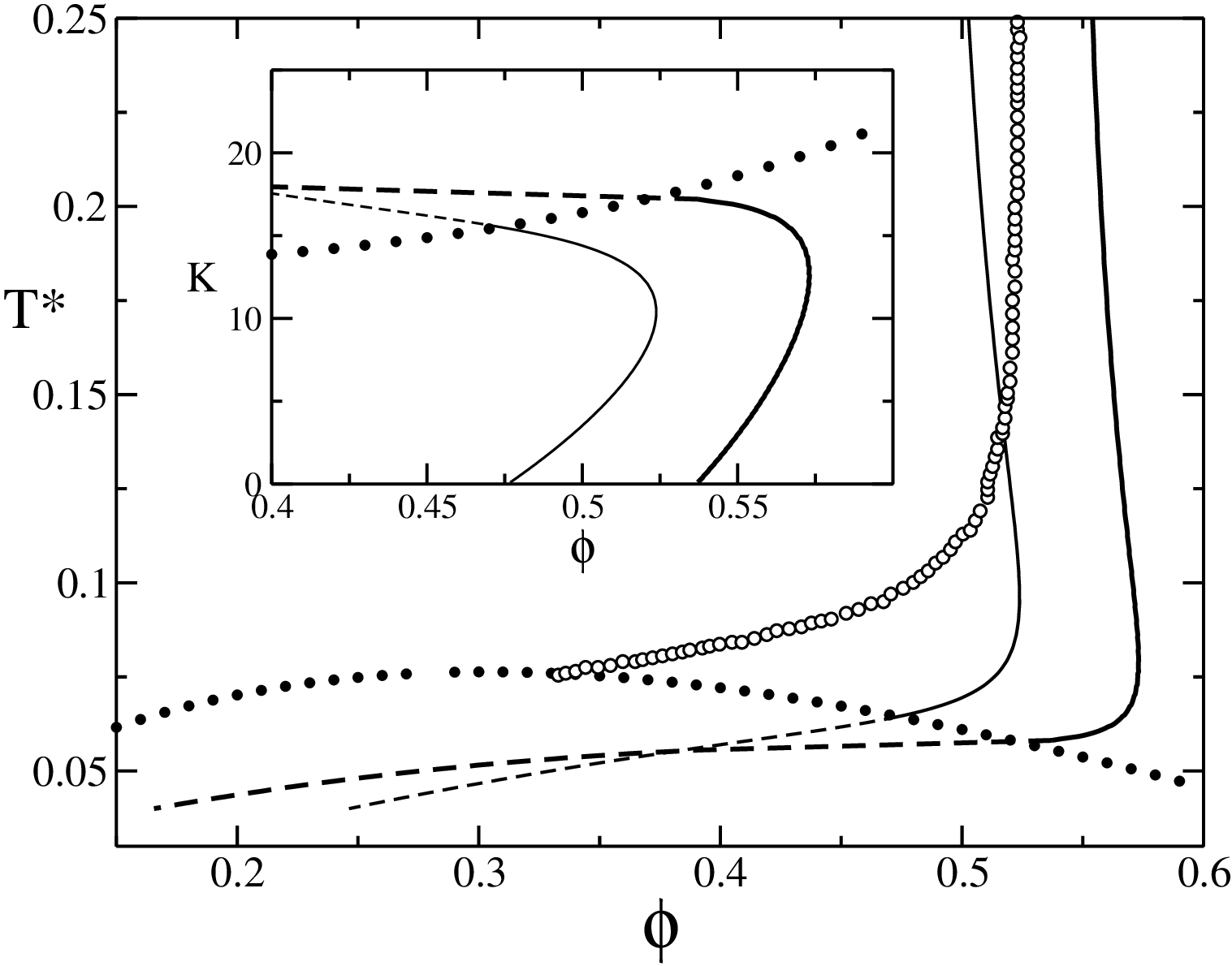}
\caption{Theoretical glass transition line in the plane ($T^*,
\phi$) for the HSAY system, calculated with Eq.\ (\ref{nep5pp})
(thick solid and dashed curve), and from the MCT
(symbols; taken from Ref.\ \cite{bergenholtz}). The lighter (solid
and dashed) line corresponds to the freezing line, according to the
HV criterium. The dotted curve corresponds to the
spinodal line. The inset displays the same information in the
($K, \phi$) plane}
\label{fig.7}
\end{center}
\end{figure}

The last illustrative example refers to another simple model of a
colloidal dispersion, this time involving the hard-sphere plus an
\textit{attractive } Yukawa (HSAY) potential, $\beta
u(r)=-K\exp[-z(r/\sigma -1)]/(r/\sigma )$, modeling depletion forces
\cite{pham}. In this case we only emphasize some qualitative aspects
of the predictions of our theory, obtained using the mean spherical
approximation (MSA) \cite{hoyeblum} for $S(k)$. Fig. 4 presents the
ergodic-non-ergodic transition line of this system for $z=20$ in the
plane ($T^*, \phi$) of state parameters, with $T^*\equiv K^{-1}$
being a reduced temperature. For reference we also plot the line
where the maximum of $S(k)$ reaches $2.85$ (the freezing line,
according to the Hansen-Verlet (HV) criterium \cite{hansenverlet}),
as well as the spinodal curve. To illustrate the qualitative
similarity between our results and those of the MCT, we have also
included the MCT transition line \cite{bergenholtz}. The inset
re-plots the same curves, now in the plane ($K, \phi$), to
illustrate the experimentally observed \cite{pham} shape of the
glass transition line at its high density end. This describes the
re-entrant behavior of a system upon lowering its temperature at
fixed volume fraction, from a hard-sphere glass to an ergodic state,
followed by the re-entrance from the ergodic state to a second
("attractive") glass state.

This completes the illustration of the practical use of the general
results in Eqs. (\ref{nep5pp}) and (\ref{nep1p}). Let us now comment
on their physical origin. Of course, the physics behind them is the
physics behind the full self-consistent theory from which they
derive. Hence, let us summarize the four distinct fundamental
elements of this theory. The first consists of general
memory-function expressions for $F(k,t)$ and $F_{S}(k,t)$ derived
with the generalized Langevin equation (GLE) formalism \cite{todos1,
5}, which in Laplace space read

\begin{equation}
F(k,z)=\frac{S(k)}{z+\frac{k^{2}D_{0}S^{-1}(k)}{1+C(k,z)}},
\label{fkz}
\end{equation}

\begin{equation}
F_{S}(k,z)=\frac{1}{z+\frac{k^{2}D_{0}}{1+C_{S}(k,z)}}, \label{fskz}
\end{equation}

\noindent where $D_0$ is the free-diffusion coefficient, and
$C(k,z)$ and $C_{S}(k,z)$ are the corresponding memory functions.

The second element is an approximate relation between collective and
self-dynamics. We approximate the difference $[C(k,t)-C_{S}(k,t)]$
by its exact short-time/large-$k$ limit, thus defining what we refer
to as the ``additive" Vineyard- like approximation \cite{5},

\begin{equation}
C(k,t)  = C_{S}(k,t) + [C^{SEXP}(k,t)- C_{S}^{SEXP}(k,t)].
\label{additive}
\end{equation}

\noindent In this equation, $C^{SEXP}(k,t)$ and $C_{S}^{SEXP}(k,t)$
are the exact short-time expressions for these memory functions,
which also define the so-called \cite{2,21} single exponential
(SEXP) approximation, and for which well-established expressions, in
terms of equilibrium structural properties, are available
\cite{todos1}.

The third ingredient consists of the independent approximate
determination of $F_{S}(k,t)$ [or $C_{S}(k,t)$]. One intuitively
expects that these $k$-dependent self-diffusion properties should be
simply related to the properties that describe the Brownian motion
of individual particles, just like in the Gaussian
approximation \cite{1}, which expresses $F_{S}(k,t)$ in terms of the
mean-squared displacement $W(t)$ as $ F_{S}(k,t)=\exp[-k^{2}W(t)]$. We
introduce an analogous approximate connection, but at the level of
their respective memory functions. The memory function of $W(t)$ is
the so-called time-dependent friction function $\Delta \zeta (t)$.
This function, normalized by the solvent friction $\zeta_0$, is the
large wave-vector limit of $C_{S}(k,t)$. Thus, we interpolate
$C_{S}(k,t)$ between its two exact limits, namely,

\begin{equation}
C_{S}(k,t)=C_{S}^{SEXP}(k,t)+\left[ \Delta \zeta^*
(t)-C_{S}^{SEXP}(k,t)\right] \lambda (k), \label{interpolation}
\end{equation}

\noindent where $\Delta \zeta^* (t)\equiv \Delta \zeta(t)/\zeta_0$.
The fourth ingredient of our theory is a general expression for this
property, also derived with the GLE approach \cite{16}, namely,

\begin{equation}
\Delta \zeta^* (t)=\frac{D_0}{3\left( 2\pi \right) ^{3}n}\int d {\bf
k}\left[\frac{ k[S(k)-1]}{S(k)}\right] ^{2}F(k,t)F_{S}(k,t).
\label{dzdt}
\end{equation}

Eqs.\ (\ref{fkz})--(\ref{dzdt}) constitute the SCGLE theory of
colloid dynamics. Besides the unknown dynamic properties, it only
involves the static structural properties $S(k)$, $C^{SEXP}(k,t)$
and $C_{S}^{SEXP}(k,t)$, determined by the methods of equilibrium
statistical thermodynamics. Concerning the interpolating function
$\lambda (k)$, phenomenological arguments were given \cite{5}  that
led to the definition given above [immediately after
Eq.\ (\ref{nep5pp})]. Although no fundamental basis is available for this
choice of $\lambda(k)$, this definition is universal (in the sense
that it is the same for any system or state), and renders the
resulting self-consistent scheme free from any form of adjustable
parameters.

The results in Eqs.\ (\ref{nep5pp}) and (\ref{nep1p}) were derived
from the long-time analysis of the SCGLE scheme in Eqs.\
(\ref{fkz})--(\ref{dzdt}). We have, however, also solved numerically
the full self-consistent theory for the systems discussed here. Of
course, we confirmed, with this lengthier method, the quantitative
results obtained in a much more economical manner from Eqs.\
(\ref{nep5pp}) and (\ref{nep1p}). The numerical solution, however,
provides the whole scenario of the relaxation processes. On the
basis of such results, which we shall discuss separately, we may say
that the general scenario of the glass transition provided by the
present theory is consistent with the available experimental data,
and is quite similar to that provided by the MCT. We should point
out, however, that for the HS system, our theory does not have to
appeal to any sort of re-scaling of the volume fraction, as the MCT
is forced to invoke due to its numerically low predicted glass
transition volume fraction ($\phi_g=0.52$). In fact, none of our
theoretical results presented here involved any form of adjustable
parameter. Thus, we conclude that in many respects, the present
theory of dynamic arrest is a sound and competitive theoretical
description of dynamic arrest in colloidal systems.

{\bf Acknowledgments:}
This work was supported by the Consejo Nacional de
Ciencia y Tecnolog\'{\i}a (CONACYT, M\'{e}xico), grants No.
C01-47611 and C02-44744. The authors are deeply indebted to Profs.
J. Bergenholtz, A. Banchio, and G. N\"agele for useful discussions.

\end{document}